\documentclass[aps,prl,twocolumn,showpacs,superscriptaddress,floatfix]{revtex4-1}
\usepackage{amsmath,amssymb}
\usepackage{graphicx}
\usepackage{epsfig}
\usepackage{color}
\usepackage{rotating}
\usepackage{bm}
\usepackage{setspace}

\begin{document}
\title{Observation of Interface Superconductivity in a SnSe$_2$-Epitaxial Graphene \\van der Waals Heterostructure}
\author{Yi-Min Zhang}
\author{Jia-Qi Fan}
\author{Wen-Lin Wang}
\affiliation{State Key Laboratory of Low-Dimensional Quantum Physics, Department of Physics, Tsinghua University, Beijing 100084, China}
\author{Ding Zhang}
\author{Lili Wang}
\author{Wei Li}
\author{Ke He}
\author{Can-Li Song}
\author{Xu-Cun Ma}
\author{Qi-Kun Xue}
\affiliation{State Key Laboratory of Low-Dimensional Quantum Physics, Department of Physics, Tsinghua University, Beijing 100084, China}
\affiliation{Collaborative Innovation Center of Quantum Matter, Beijing 100084, China}

\begin{abstract}
We report on the direct observation of interface superconductivity in single-unit-cell SnSe$_2$ films grown on graphitized SiC(0001) substrate by means of van der Waals epitaxy. Tunneling spectrum in the superconducting state reveals rather conventional character with a fully gapped order parameter. The occurrence of superconductivity is further confirmed by the presence of vortices under external magnetic field. Through interface engineering, we unravel the mechanism of superconductivity that originates from a two-dimensional electron gas formed at the interface of SnSe$_2$ and graphene. Our finding opens up novel strategies to hunt for and understand interface superconductivity based on van der Waals heterostructures.

\end{abstract}

\maketitle
\begin{spacing}{1}
Interface superconductivity has recently been the subject of numerous studies for the condensed matter community \cite{qing2012interface, reyren2007superconducting, gozar2008high, he2014two, zhong2016nodeless, saito2017highly}. This appears to be understandable from the perspective of fundamental research since the superconductivity confined in a two-dimensional (2D) interface exhibits many exotic phenomena that have certain counterparts in layered cuprates and iron-based superconductors \cite{richter2013interface, saito2015metallic, saito2017highly}, and thus providing unprecedented opportunities to crack the mystery of high temperature ($T_\textrm{c}$) superconductivity therein. It seems more significant insofar as the superconducting technology application is concerned. By constructing and tailoring hybrid heterostructures, the interface might benefit from the two building blocks and exhibit an unexpectedly high $T_\textrm{c}$ \cite{qing2012interface, ge2014superconductivity}. Moreover, the modified fluctuations, electron correlations and spin-orbit coupling in reduced dimensions are potential factors to drive the emergence of novel quantum phenomena \cite{he2014two, gong2017time}, paving the unique way to pursue more promising technologies. Despite extensive research efforts, however, a unified microscopic picture on how the interface superconductivity occurs remains as enigmatic as ever \cite{wu2013anomalous, huang2017monolayer, brun2016review}, in part due to the complexity of interface involved. It is therefore highly tempting to buildup much simpler superconducting heterostructures.

Tin diselenide (SnSe$_2$), a main-group metal dichalcogenide and being superconducting by organometallic intercalation \cite{formstone1990observation, anthonyacox1991single, o1992relatively, li2017molecule}, exhibits the similar layered structure with graphene and transition metal dichalcogenides (TMDCs). Recent extensive studies have revealed rich physics and potential applications in these materials \cite{bhimanapati2015recent}. For example, superconducting and electrically gated TMDCs not only show many properties looking analogous to those observed in cuprates \cite{klemm2015pristine}, but also present new electron pairing with nontrivial topology, such as the the 2D Ising superconductivity protected by spin-valley locking \cite{lu2015evidence, saito2016superconductivity, zeng2018gate}. In this study, we grow high-quality SnSe$_2$ films on graphitized SiC(0001) substrate, and present unambiguous evidence of superconductivity at the van der Waals (vdWs) interface of SnSe$_2$ and graphene by using scanning tunneling microscopy (STM). By exploring the variances of film thickness and graphene, we tailor the hybrid hetrostructrue and reveal a 2D electron gas (2DEG) formed at the SnSe$_2$ and graphene interface, which bears the responsibility for superconductivity observed there.

Our experiments are carried out on an ultrahigh vacuum cryogenic STM system (Unisoku) equipped with a molecular beam epitaxy (MBE) for sample preparation. The base pressure of both chambers is better than 1.0 $\times$ 10$^{-10}$ Torr. Nitrogen-doped SiC(0001) wafers (0.1 $\Omega\cdot$cm) are graphitized by being heated to 1350$^\circ$C, which results in a bilayer graphene-dominant surface \cite{hass2008growth}. High purity Sn (99.9999$\%$) and Se (99.999$\%$) sources are co-evaporated from standard effusion cells (CreaTec) on the graphitized SiC(0001) substrate at about 210$^\circ$C, giving rise to a layer-by-layer epitaxial growth of SnSe$_2$ films. A higher substrate temperature of 240$^\circ$C leads to a transition from SnSe$_2$ to the cubic SnSe phase. During the MBE growth, a Se-rich atmosphere is used to compensate for the loss of volatile Se, bearing a similar growth dynamics with the one for other metal selenides \cite{song2011molecular}. Once the film growth is stopped, the SnSe$_2$/graphene heterostructures are \textit{in-situ} transferred into the STM head for data collection. A polycrystalline PtIr tip, cleaned by electron beam heating and calibrated on epitaxial Ag/Si(111) films, is used throughout the experiments. Tunneling spectra and maps are measured at 0.4 K by using a standard lock-in technique with a small bias modulation of 0.1 meV at 931 Hz, unless other specified.
\end{spacing}

\begin{figure}[t]
\includegraphics[width=\columnwidth]{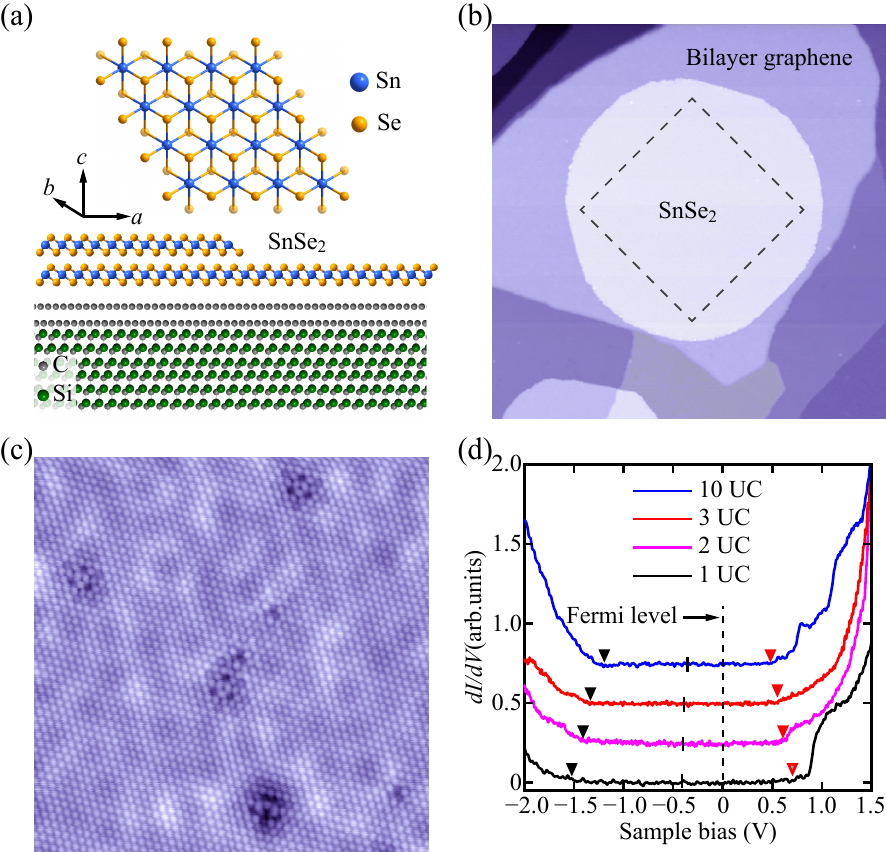}
\caption{(color online) (a) Sketch of SnSe$_2$/graphene vdWs heterostructure. (b) Typical topographic image (300 nm $\times$ 300 nm, $V$ = 3.5 V, $I$ = 20 pA) of \textit{in-situ} grown SnSe$_2$ films on graphitized SiC(0001). The dashed square marks the region where ZBC maps are acquired later (c) Zoom-in STM topography (18 nm $\times$ 18 nm, $V$ = 50 mV, $I$ = 100 pA) on 1-UC SnSe$_2$ film. The bright spots denote the Se atoms at the top layer. (d) Thickness-dependent \textit{dI/dV} spectra of SnSe$_2$ films. The black and red triangles mark respectively the VBM and CBM of SnSe$_2$, with their middles represented by the short vertical lines. Tunneling gap is set at $V$ = 1.5 V and $I$ = 150 pA. The lock-in bias modulation has a magnitude of 10 meV.
}
\end{figure}

As a layered semiconductor, SnSe$_2$ crystallizes into the CdI$_2$ type structure and consists of a hexagonally packed layer of Sn atoms sandwiched between two layers of Se anions \cite{zeng2018gate, Fong1972electronic}. The intralayer Sn-Se bonds are predominantly covalent in nature, whereas the forces between the sandwich layers are of weak vdWs type. Figure 1(a) schematically draws the geometry of epitaxial SnSe$_2$ films on graphitized SiC(0001) substrate, in which the middle bilayer graphene and the top SnSe$_2$ films are stacked by weak vdWs interactions. Figure 1(b) typifies a constant-current STM topographic image of as-grown SnSe$_2$ films, with a nominal thickness of about 0.7 unit cell (UC, one Se-Sn-Se triple layer). A magnified STM image reveals the top Se atoms, which are in a hexagonal close packing and spaced $\sim$ 3.82 $\pm$ 0.03 $\textrm{\AA}$ apart [Fig.\ 1(c)]. This value, together with the extracted out-of-plane lattice constant of approximately 6.1 $\textrm{\AA}$ by measuring the height difference across the steps of SnSe$_2$ epitaxial films, match excellently with the lattice parameters for SnSe$_2$ \cite{Fong1972electronic}. Furthermore, we carry out the film-thickness-dependent analysis and find no observable variation in the lattice constants. However, the electronic band structures vary significantly with the film thickness, as clearly revealed in Fig.\ 1(d). As the film reduces in thickness, both the valance and conduction bands of SnSe$_2$ films move away from the Fermi level ($E_F$), leading to an obvious increase in the band gap E$_\textrm{g}$. This can be quantatively seen in the Supplemental Fig.\ S1 \cite{supplementary}, in which we measure the energy positions of valance band maximum (VBM) and conduction band minimum (CBM) of SnSe$_2$ as well as calculate the direct band gap E$_\textrm{g}^{\textrm{dir}}$. The increased E$_\textrm{g}^{\textrm{dir}}$ has been theoretically attributed to the poor electrostatic screening and enhanced quantum confinement of electrons in few-layer SnSe$_2$ systems \cite{Gonzalez2016layer}. Despite the variations, the halfway between VBM and CBM changes little with the film thickness and is pinned at $\sim$ 0.4 eV below $E_F$.

\begin{figure}[t]
\includegraphics[width=1\columnwidth]{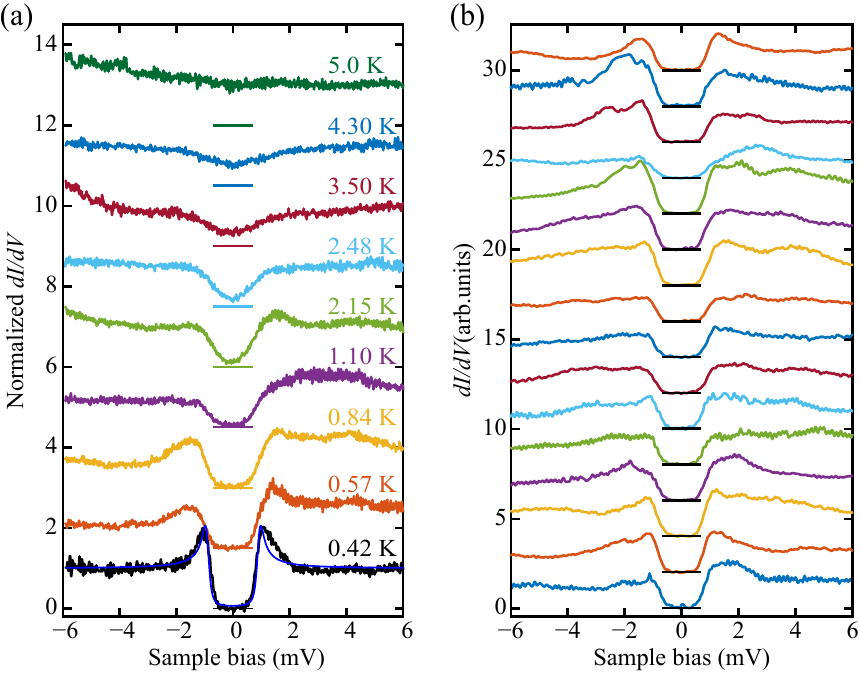}
\caption{(color online) (a) Tunneling \textit{dI/dV} spectra ($V$ = 6 mV, $I$ = 100 pA) of 1-UC SnSe$_2$ film on bilayer graphene as a function of temperature as indicated. For clarity the spectra have been vertically offset, with their zero conductance positions marked by correspondingly colored horizontal lines. The same convention is used throughout, unless otherwise noted. The blue line at bottom shows the best fit of experimental data (black curve) to the BCS Dynes formula with a single isotropic $s$-wave gap. (b) A series of \textit{dI/dV} spectra ($V$ = 10 mV, $I$ = 100 pA) acquired at equal separation along a 40 nm-trajectory, revealing substantial inhomogeneity of the superconducting gap at 0.4 K. Instead of colored lines, the black lines mark the zero conductance positions for clarity.
}
\end{figure}

Strikingly, in the large semiconducting band gap high-resolution tunneling spectroscopy within a narrower energy range of $\pm$ 6 meV at 0.4 K, as shown by the black curve in Fig.\ 2(a), discloses an $E_F$-symmetric and fully-gapped density of states (DOS), which we ascribe as the superconducting energy gap in SnSe$_2$/graphene heterostructure. In some regions ($\sim$ 25\%), the superconducting energy gap has pronounced coherence peaks and can be reasonably described by the well-known BCS Dynes expression with a single isotropic \textit{s}-wave gap and adjustable lifetime broadening \cite{Dynes1978direct}. A representative fit to such a spectrum in Fig.\ 2(a) yields an energy gap with magnitude $\Delta$ = 0.95 meV (blue line). The temperature dependence of the tunneling spectra shows the progressive suppression of superconducting coherence peaks and lift of zero bias conductance (ZBC) at elevated temperature [Fig.\ 2(a)], with the behaviors in excellent agreement with superconductivity. The superconducting gap eventually vanishes at temperatures close to a transition temperature $T_\textrm{c}$ of 4.84 K [Fig.\ S2(a)]. It is worth noting that the superconducting gap critically depends on the position of the STM tip on the SnSe$_2$ films and exhibits significant spatial inhomogeneity [Fig.\ 2(b)], which might most likely originate from the structural ripples of epitaxial graphene related to the $6\sqrt{3} \times 6\sqrt{3}$ reconstruction on SiC(0001) \cite{hass2008growth}. This is evident from atomically-resolved STM images [Fig.\ 1(c)], in which the underlying graphene/SiC superstructrure can be clearly seen. Albeit with the site-dependent fine structure, we notice that there always exists a vanishing DOS over a finite energy range near $E_F$ in the \textit{dI/dV} spectra, suggesting a rather conventional character of the superconductivity with a fully gapped order parameter. A statistical study of gap magnitude $\Delta$ [Fig.\ S1(b)], defined as half the energy separation between the two gap edges, reveals a predominant distribution of $\Delta$ close to 1.05 meV. This somewhat overestimates the $\Delta$ and results in an upper limit of the reduced gap ratio $2\Delta/k_\textrm{B}T_\textrm{c}\sim$ 5.04. The superconductivity in the SnSe$_2$/graphene heterostructure created here might be a strong-coupling type.

In order to further confirm the superconductivity, we have carried out the tunneling experiments under a varying magnetic field. Application of the field perpendicular to superconducting SnSe$_2$/graphene vdWs heterostructure can locally kill the superconductivity and lead to the appearance of Abrikosov vortices, each carrying a quantized flux \textit{h}/2e. To search for such vortices, we map out the spatial-resolved ZBC (64 pixels $\times$  64 pixels) under various magnetic field on a 120 nm $\times$ 120 nm field of view of SnSe$_2$ films, outlined in Fig.\ 1(b). Figures 3(a) and 3(b) present the ZBC maps, in which the yellow regions with enhanced ZBC signify the penetration of vortices into the heterostructrue. Although three individual isolated vortices are expected and actually identified at 0.5 T [Fig.\ 3(a)], at a higher field of 1.0 T the vortices get close to each other and cluster into the central field of view [Fig.\ 3(b)]. Note that the irregular vortex core might be due to the inhomogeneous superconducting state in SnSe$_2$/graphene heterostructure [Fig.\ 2(b)], bearing a striking resemblance to cuprate superconductors having the notorious electronic inhomogeneity \cite{Fischer2007scanning}. As plotted in Fig.\ 3(c) are a series of \textit{dI/dV} spectra taken at equal separations (7.5 nm) across a vortex core in Fig.\ 3(a). Evidently, the spatial dependence of such tunneling conductance spectra reveals the disappearance of the superconducting gap at sites close to the vortex center (black curves). No quasiparticle bound state is found within the vortex core, primarily due to the graphene ripple-induced electron scattering \cite{katsnelson2008electron}, which reduces the electron mean free path and pushes the superconducting SnSe$_2$/graphene into the dirty limit \cite {Renner1991scanning}. In any cases, our direct visualization of vortices has established unambiguous evidence of superconductivity in SnSe$_2$/graphene vdWs heterostructure.

\begin{figure}[t]
\includegraphics[width=0.96\columnwidth]{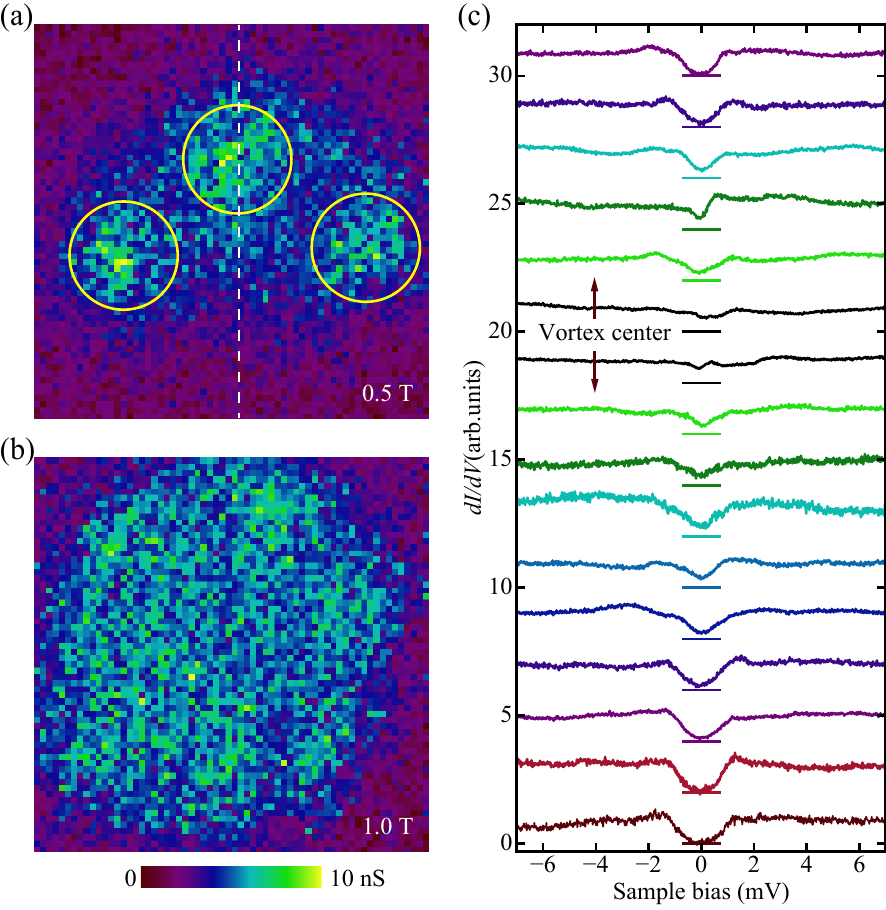}
\caption{(color online) (a) ZBC map (120 nm $\times$ 120 nm) showing three individual isolated vortices (emphasized by yellow circles) of 1-UC SnSe$_2$/graphene heterostructure at 0.5 T. The white dashed line designates the trajectory along which the \textit{dI/dV }spectra in (c) are measured. The tunneling junction (also applies to (b) and (c)) is set at $V$ = 8 mV and $I$ = 100 pA. (b) Vortex clustering mapped in the same field of view as (a) and at a higher magnetic field of 1.0 T. (c) Linecut of \textit{dI/dV} spectra taken through a vortex core in (a).
}
\end{figure}

\begin{figure*}[t]
\includegraphics[width=1.77\columnwidth]{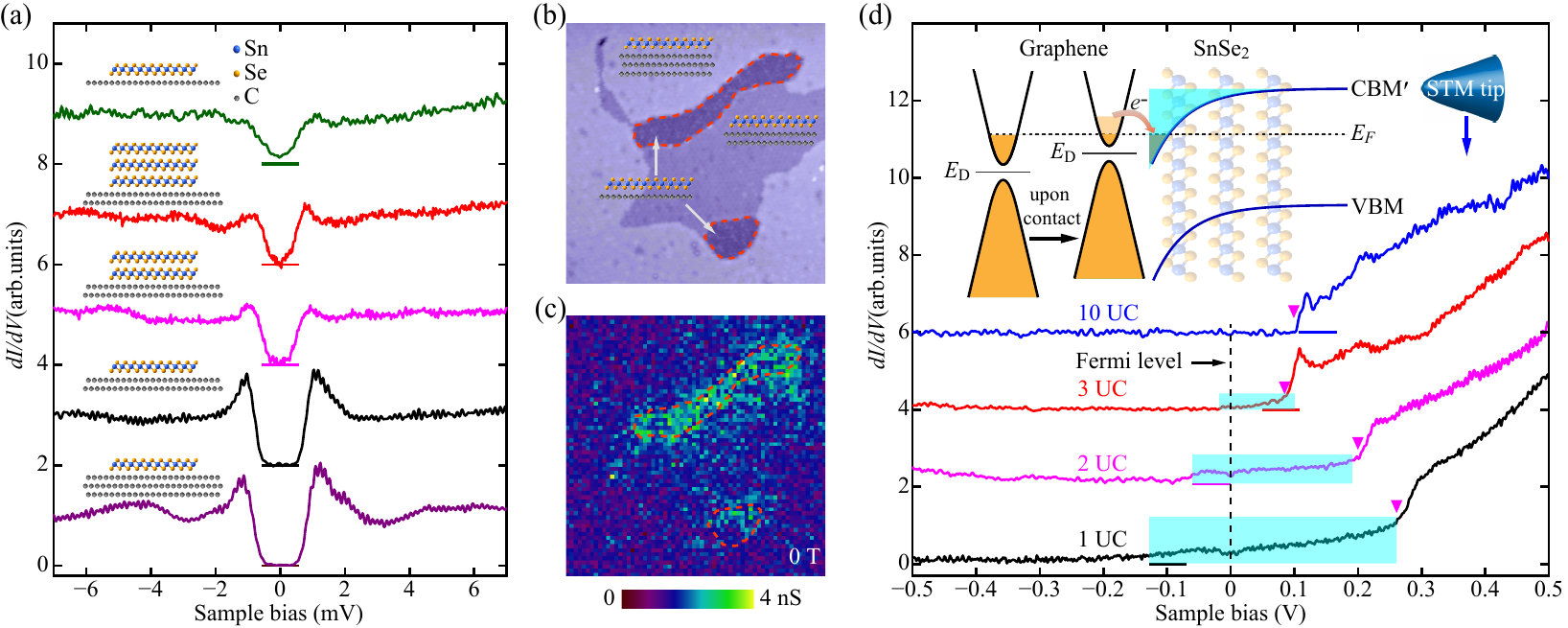}
\caption{(color online) (a) Tunneling spectra \textit{versus} the sketched SnSe$_2$/graphene vdWs heterostructure on the upper left side of every curve. For simplicity the atoms of SiC are not displayed. The tunneling junction is set at $V$ = 10 mV and $I$ = 100 pA except for the bottom one ($V$ = 8 mV, $I$ = 100 pA). (b) A 98 nm $\times$ 98 nm STM topographic image ($V$ = 10 mV, $I$ = 100 pA) of 1-UC SnSe$_2$ films prepared on monolayer, bilayer and trilayer grpaphene coexisting substrate. The red dashes encircle the region of SnSe$_2$ films situated on monolayer graphene. (c) Simultaneous ZBC map (64 pixels $\times$  64 pixels) revealing graphene layer-dependent conductance contrast at zero field. (d) Dependence of $dI/dV$ spectra ($V$ = 0.5 V, $I$ = 100 pA) on the SnSe$_2$ film thickness, measured in an intermediate energy range of $\pm$ 0.5 eV. The lock-in has a bias modulation of 5 meV. The magenta triangles mark the CBM$^\prime$ at the \textit{M} point of SnSe$_2$, located far above $E_F$ (vertical dashes). Inset shows the energy band scheme for SnSe$_2$/graphene vdWs heterostructure.
Electron transfer from graphene to the SnSe$_2$ films is indicated by the orange arrow. The band gap opening near the Dirac point ($E_D$) of graphene due to the SiC substrate is also shown.
}
\end{figure*}

In what follows, we engineer the SnSe$_2$/graphene vdWs heterostructure and shed light on the mechanism of superconductivity. It is worth pointing out that in addition to bilayer graphene, monolayer and trilayer graphene are also available on the graphitized SiC(0001) wafers \cite{hass2008growth}. Meanwhile, SnSe$_2$ films with any desired thickness can be obtained by controlling the growth duration. Both of the variances provide us with unique opportunities to tailor the SnSe$_2$/graphene vdWs heterostructure and clarify their roles in superconductivity. Enumerated in Fig.\ 4(a) are the two key results we disclose. First, the superconducting energy gap reduces in magnitude $\Delta$ and gets filled with subgap DOS as the SnSe$_2$ film thickness is increased (cf.\ the three curves in the middle of Fig.\ 4(a)). This indicates a suppressed superconductivity with increasing SnSe$_2$ film thicknes, and provides the first piece of evidence that the superconductivity might occur at the interface of SnSe$_2$ and graphene. Indeed, tunneling spectrum on thicker films reveals a semiconducting character and the STM tip can be never stabilized at voltages smaller than 0.1 V. Second, the number of graphene layer plays a vital role in the superconducting state. Although the bilayer and trilayer graphene give rise to U-shaped pairing gap, the heterostructures composed of SnSe$_2$ and monolayer graphene are typically sized of V-shaped gaps with nonzero subgap DOS at $E_F$ and no coherence peak (green curve). To reveal that this does not happen by accident, we map out the zero-field spatial ZBC of SnSe$_2$ grown on a substrate region of coexisting monolayer, bilayer and trilayer graphene [Fig.\ 4(b)]. As confirmed in Fig.\ 4(c), the SnSe$_2$ films situated on monolayer graphene universally exhibit enhanced ZBC and weak superconductivity. This is consistent with the preferential vortex pinning at locations of the SnSe$_2$/monolayer graphene heterostructure [Fig.\ S3].

In SnSe$_2$/graphene vdWs heterostructure we construct here, a simple explanation of superconductivity by either strain-induced lattice distortion or element interdiffusion seems unlikely. For vdWs epitaxy, strains are often small, and if they came into play the superconductivity should not rely significantly on the number of graphene layer as observed in Fig.\ 4(a). Moreover, we reveal no signature of superconductivity on uncovered graphene after the SnSe$_2$ growth, which rules out a possible cause of superconductivity by diffusion of Sn and/or Se into graphene. On the other hand, an inverse diffusion of carbon into SnSe$_2$ film is nearly impossible as well if the low growth temperature of 210$^\circ$C is considered. Given that the superconductivity is sharply dependent on the two materials building the heterostructure [Fig.\ 4(a)], a more plausible explanation would seem to be that the superconductivity stems from the interface between SnSe$_2$ and graphene. Learning that epitaxial graphene on SiC has a lower work function (4.2 eV$ \sim$ 4.4 eV) \cite{mammadov2017work} than SnSe$_2$ ($\sim$ 5.3 eV) \cite{shimada1994work}, upon contact electrons would flow from graphene to SnSe$_2$, leading to a downwards band bending of the SnSe$_2$ bands and electron accumulation near the interface. At equilibrium, their Fermi levels are aligned and a 2DEG is created at the SnSe$_2$/graphene interface. This is clearly illustrated in the inserted energy band diagram of Fig.\ 4(d), and supported by our experiments. As plotted in Fig.\ 4(d) are the thickness-dependent \textit{dI/dV} spectra recorded in an intermediate energy region of $\pm$ 0.5 eV, from which two findings are immediately revealed. First, each \textit{dI/dV} spectrum presents a prominent drop of DOS around 0.1$\sim$0.3 eV, which we interpret as DOS variation from the conduction band at the \textit{M} high-symmetry point of SnSe$_2$ \cite{Gonzalez2016layer}. This allows for determination of its minimum (dubbed as CBM$^\prime$) and indirect band gap $\textrm{E}_\textrm{g}^{\textrm{in}}$ of SnSe$_2$ [Fig.\ S1], which shows a quantatative agreement with the theoretical calculations \cite{Gonzalez2016layer}. Second, and the most significantly, the band edges get rounded and show an increasingly long nonzero DOS tail (cyan-marked) toward $E_F$ with reduced film thickness. A closer inspection of the band scheme in Fig.\ 4(d) reveals immediately its origin from the 2DEG confined at the SnSe$_2$/graphene interface. Note that all bulk bands of SnSe$_2$ are positioned far away from $E_F$ (e.g.\ $>$ 0.2 eV for 1-UC SnSe$_2$) and the nonzero DOS in thin films originate solely from 2DEG, we argue that the superconductivity occurs due to the formation of 2DEG in SnSe$_2$/graphene heterostructure. Indeed, the nonzero DOS and 2DEGs get shrinking with the film thickness, matching well with the suppressed superconductivity in thick SnSe$_2$ films.

Notably, the graphene layer-dependent superconductivity seems counterintuitive since monolayer graphene has the smallest work function \cite{mammadov2017work} and is more beneficial to the 2DEG formation and superconductivity. However, one should be aware that the stronger ripples of monolayer graphene will certainly cause strong electron scattering \cite{hass2008growth} that is harmful to superconductivity. Besides, a more relevant factor may be the relatively lower carrier density in monolayer graphene as compared to bilayer and trilayer graphene \cite{Ohta2007interlayer}. This leads to a 2DEG with low concentration, which, in conjunction with the strong scattering, may be responsible for the V-shaped gap structure and suppressed superconductivity. A recent study of SnSe$_2$ bilayer prepared on graphite also revealed V-shaped spectral gaps, but with an unreasonably large gap size $\Delta$ of $\sim$ 16-22 meV as compared to $T_c$, which were interpreted as a signature of unconventional superconductivity \cite{Mao2018signature}. However, without vortex imaging and careful interface engineering explored here, whether the large gap relates to superconductivity is highly doubtful.

Our detailed STM scrutiny of SnSe$_2$/graphene vdWs heterostructure has discovered clear interface superconductivity with a rather conventional character. The revealed mechanism of superconductivity due to the formation of 2DEG might shed important insight into interface superconductivity as well as the mechanism of high-$T_c$ superconductivity in compounds made of many heterostructures at the atomic plane limit. Moreover, our study suggests that the semiconducting SnSe$_2$ and its heterostructures can serve as ideal platforms to explore the physics of interface superconductivity. Further interface engineering through preparing SnSe$_2$ on substrates with high carrier densities and electron-phonon coupling (e.g.\ perovskite oxide SrTiO$_3$) might promote superconductivity with a higher critical temperature.

\begin{acknowledgments}
This work is financially supported by the Ministry of Science and Technology of China (Grants No. 2017YFA0304600, 2016YFA0301004) and the National Natural Science Foundation of China (Grants No. 11427903, 11504196, 11634007, 11774192). C. L. S. acknowledges support from the National Thousand-Young-Talents Program and Tsinghua University Initiative Scientific Research Program.
\end{acknowledgments}

%

\widetext
\setcounter{equation}{0}
\setcounter{figure}{0}
\setcounter{table}{0}
\makeatletter
\renewcommand{\theequation}{S\arabic{equation}}
\renewcommand{\thefigure}{S\arabic{figure}}
\renewcommand{\bibnumfmt}[1]{[S#1]}
\renewcommand{\citenumfont}[1]{S#1}

\onecolumngrid
\Large
{\textbf{Supplemental Material for:} }
\begin{center}
\textbf{\large Observation of Interface Superconductivity in a SnSe$_2$-Epitaxial Graphene \\van der Waals Heterostructure}
\end{center}

\small
\maketitle
\onecolumngrid
\begin{figure}[h]
\includegraphics[width=0.6\columnwidth]{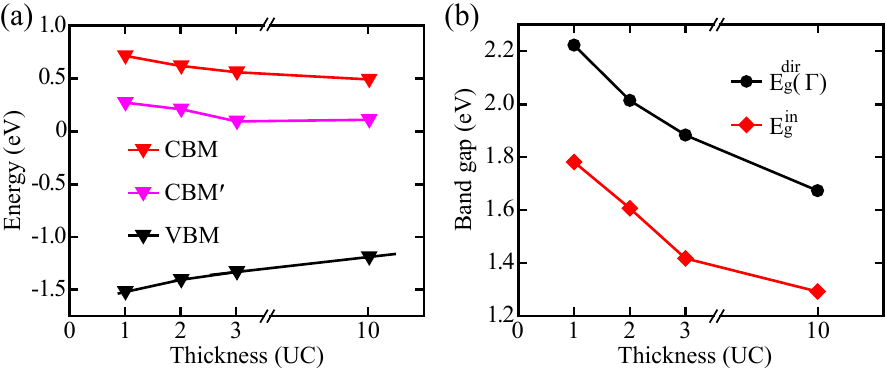}
\caption{ (a) Energy positions of VBM, CBM and CBM$^\prime$ as a function of the SnSe$_2$ film thickness. The abbreviations of CBM and CBM$^\prime$ respectively correspond to the conduction band minima at the $\Gamma$ and M high-symmetry points of SnSe$_2$, while the VBM the valance band maximum at the $\Gamma$ point. (b) Thickness-dependent semiconducting band gaps for SnSe$_2$ epitaxial films on graphene substrate. Note that the direct band gap E$_\textrm{g}^{\textrm{dir}}(\Gamma)$ is measured to be the difference between VBM and VBM at the $\Gamma$ point of SnSe$_2$, which is larger than the actual one taking place at the M point.
}
\end{figure}

\begin{figure}[h]
\includegraphics[width=0.57\columnwidth]{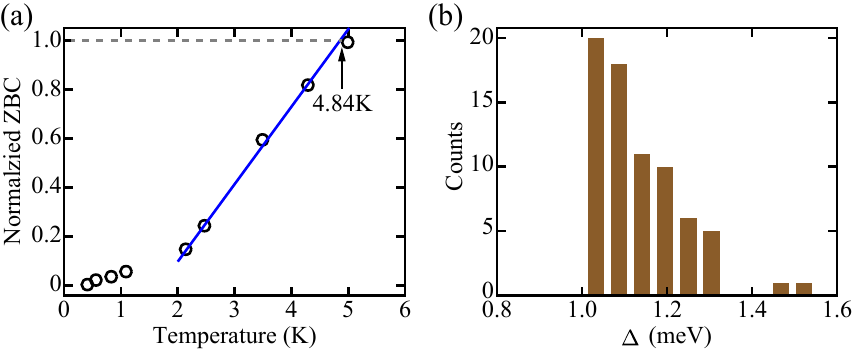}
\caption{(a) Normalized ZBC as a function of temperature, showing a linear dependence between them at temperatures close to $T_\textrm{c}$. By extrapolating $T_\textrm{c}$ to the point where ZBC = 1, the $T_\textrm{c}$ of 4.84 K is revealed in the vdWs heterostructrue of 1-UC SnSe$_2$ and bilayer graphene. (b) Statistics of superconducting gap magnitude $\Delta$.
}
\end{figure}

\begin{figure}[h]
\includegraphics[width=0.6\columnwidth]{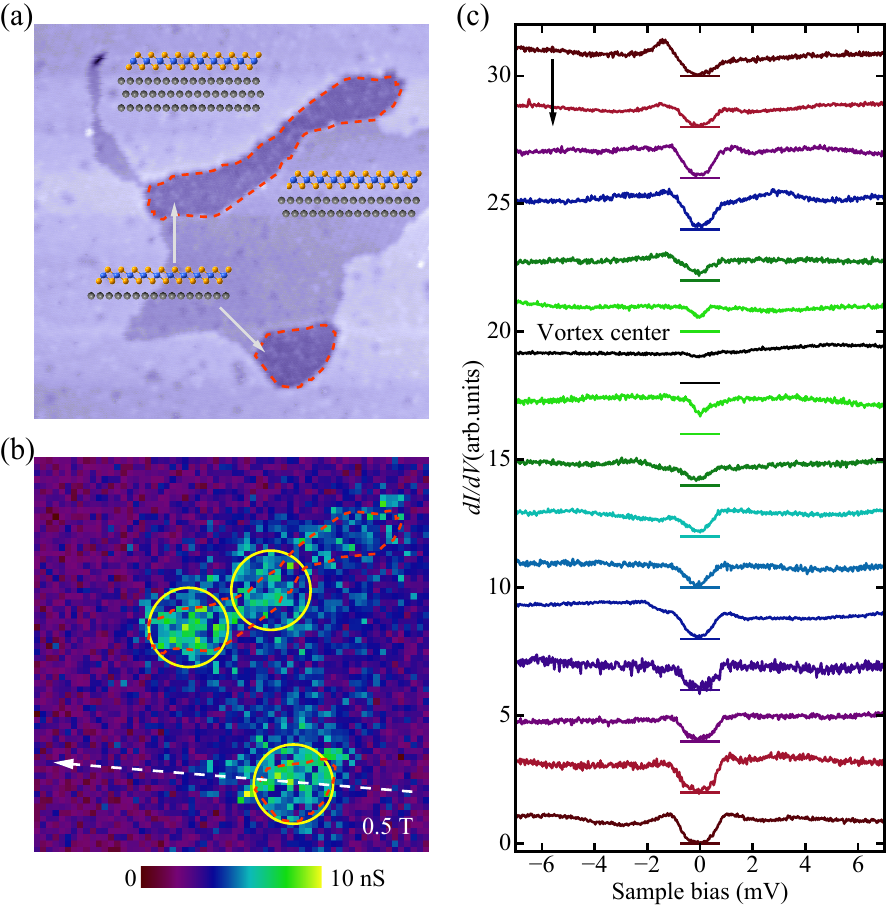}
\caption{(a) STM topographic image (98 nm $\times$ 98 nm, $V$ = 10 mV, $I$ = 100 pA) of 1-UC SnSe$_2$ films prepared on monolayer, bilayer and trilayer grpaphene coexisting substrate. The red dashes encircle the region of 1-UC SnSe$_2$ films situated on monolayer graphene. (b) Simultaneous ZBC map taken at 0.5 T, showing the preferential vortex pining at the red dashes-encircled region. The dashed arrow indicates the trajectory along which the \textit{dI/dV} spectra in (c) are measured. (c) Line-cut \textit{dI/dV} spectra across one vortex core in the right bottom corner of (b), revealing the disappearance of superconductivity near the vortex center.
}
\end{figure}

\end{document}